\documentclass[sigconf]{acmart}

\AtBeginDocument{%
  }

\copyrightyear{2024}
\acmYear{2024}
\setcopyright{rightsretained}
\acmDOI{10.1145/3627673.3679077}

\acmConference[CIKM '24] {Proceedings of the 33rd ACM International Conference on Information and Knowledge Management}{October 21--25, 2024}{Boise, ID, USA.}
\acmISBN{979-8-4007-0436-9/24/10}

\begin{document}

\title{Advertiser Content Understanding via LLMs for Google Ads Safety}


\author{Joseph Wallace}
\email{joseywallace@google.com}
\affiliation{%
  \institution{Google Ads Safety}
  \city{Mountain View}
  \country{USA}
}

\author{Tushar Dogra}
\email{tdogra@google.com}
\affiliation{%
  \institution{Google Ads Safety}
  \city{Mountain View}
  \country{USA}
}

\author{Wei Qiao}
\email{weiqiao@google.com}
\affiliation{%
  \institution{Google Ads Safety}
  \city{Mountain View}
  \country{USA}
}

\author{Yuan Wang}
\email{yuanwang@google.com}
\affiliation{%
  \institution{Google Ads Safety}
  \city{Mountain View}
  \country{USA}
}

\renewcommand{\shortauthors}{Wallace et al.}

\begin{abstract}
Ads Content Safety at Google requires classifying billions of ads for Google Ads content policies. Consistent and accurate policy enforcement is important for advertiser experience and user safety and it is a challenging problem, so there is a lot of value for improving it for advertisers and users. Inconsistent policy enforcement causes increased policy friction and poor experience with good advertisers, and bad advertisers exploit the inconsistency by creating multiple similar ads in the hope that some will get through our defenses. This study proposes a method to understand advertiser's intent for content policy violations, using Large Language Models (LLMs). We focus on identifying good advertisers to reduce content over-flagging and improve advertiser experience, though the approach can easily be extended to classify bad advertisers too. We generate advertiser's content profile based on multiple signals from their ads, domains, targeting info, etc. We then use LLMs to classify the advertiser content profile, along with relying on any knowledge the LLM has of the advertiser, their products or brand, to understand whether they are likely to violate a certain policy or not. After minimal prompt tuning our method was able to reach 95\% accuracy on a small test set.

\end{abstract}




\received{27 June 2024}
\received[accepted]{16 July 2024}

\maketitle

\section{Problem and Motivation}

Advertising at Google aims to enhance user experience by displaying relevant and high-quality ads. However, maintaining ad quality at Google's massive scale presents a significant challenge. With billions of ads served daily, ensuring each ad adheres to Google's content policies necessitates highly efficient and automated solutions. Even a small percentage of misclassifications can damage user trust and harm Google's brand.

Traditional machine learning approaches, primarily focused on classifying individual ads~\cite{scaling_up_llm}, often lack the context necessary for accurate classification. A human reviewer benefits from understanding the advertiser's brand and overall messaging, recognizing that a seemingly innocuous phrase might be problematic within a specific context. However, manual human review of every advertiser is infeasible due to the vast volume of advertisers each of which may have thousands of individual ads.

This paper introduces a novel approach to ad content policy classification leveraging the power of long-context Large Language Models (LLMs).  Our technique shifts from individual ad classification to an advertiser-centric approach. By processing vast amounts of data related to an advertiser, including their website content, news mentions, and social media presence, the LLM develops a comprehensive understanding of the brand. This contextual knowledge, combined with the LLM's analysis of the advertiser's ads, enables more nuanced and accurate classification decisions. We focus our evaluation on the "Non-Family Safe" (NFS) policy, a category prone to false positives, demonstrating how our LLM-driven approach significantly improves accuracy and reduces misclassifications.

\begin{figure}[t]
  \centering
  \includegraphics[width=\linewidth]{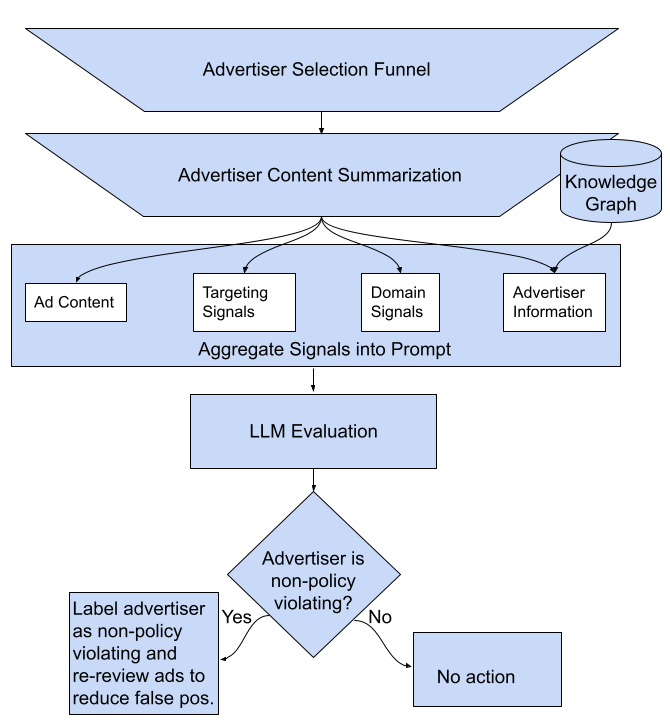}
  \caption{A diagram of our end-to-end solution for Advertiser Content Understanding showing the funneling of advertisers and content summarization from which various features are aggregated and prepared into a prompt for LLM evaluation.}
  \Description{An end-to-end solution for advertiser content understanding via LLMs for ads content safety.}
  \label{fig:acu}
\end{figure}

\section{Method}
At a high level, our approach has multiple steps as shown in Fig.~\ref{fig:acu}. The first step is to select advertisers that we should process via LLMs. The second step is to generate the content summary for those advertisers. The third step is to prepare the LLM prompt based on the content summary and the advertiser knowledge. The final step is LLM classification and storing the results. All LLM evaluations were performed using Gemini 1.0 Ultra~\cite{geminiteam2024gemini}. We describe each of these steps in detail below.

\subsection{Advertiser Candidate Selection}
Advertisers are selected for LLM evaluation based on the degree to which their ads are potentially over-flagged by baseline, ad level classifiers, since running LLMs on all advertisers is prohibitively expensive.

\subsection{Advertiser Content Summarization}
Similarly, to advertiser selection above, an individual advertiser may have thousands of ads which would be expensive and inefficient to evaluate in its entirety. Instead, we filter their content (as shown in Fig.~\ref{fig:acu}) to either the most relevant content, already labeled content, and known false positives from the advertiser. Together these criteria provide a holistic summary of the advertiser's content with respect to the policy of interest. 

\subsection{Feature Aggregation}
The summarized content above includes ad content, and other signals like targeting information, domain signals, as well as advertiser information from Knowledge Graph~\cite{knowledge_graph} (if available). These features are aggregated and deduplicated to form a content profile of the advertiser.

\subsection{Prompt Engineering}
The advertiser content profile is paired with a description of the Google ads Non-Family Safe policy for which the advertiser is being evaluated and the LLM is prompted to (1) summarize the advertiser and their content, (2) describe the advertiser’s products and services, and (3) determine if the advertiser is violating the Google ad policy. The Google ads policy includes examples of content that is in-scope and out-of-scope of the policy.

For prompt tuning we hand-labeled several hundred advertisers and split them into three groups. The first two were used to iteratively improve the prompt and the third dataset was held for final validation of prompt tuning.

To improve the prompt, the LLM was run on a subset of the hand-labeled examples. Each incorrect classification was binned into categories describing the type of error. This allowed us to identify areas of the prompt that were under-constrained and improve. After only two iterations we were able to substantially improve performance. 

\section{Results}

% \subsection{Model Performance Metrics}
We ran this method for the Non-Family Safe policy, and the table below shows the precision, recall, and accuracy metrics for identifying good advertisers after two rounds of prompt tuning. 

The precision, recall, and accuracy of the LLM are computed on a small sample of manually reviewed advertisers. The recall and precision are defined in the standard method. Recall is equal to the percentage of truly non-policy violating advertisers that the LLM identifies from the entire dataset.  Precision is equal to the percentage of truly non-policy violating advertisers from those the LLM identifies as non-policy violating.

\vspace{\baselineskip}

\begin{center} 
\begin{tabular}{ |c|c|c| }  
 \hline 
 Accuracy & Precision & Recall \\  
 95\% & 97\% & 95\% \\   
 \hline 
\end{tabular} 
\end{center} 

\vspace{\baselineskip}

The metrics are quite impressive given the difficulty of holistic classification and that the model has no prior training on this specific task and should continue to improve as (1) the LLM models improve and (2) with further prompt tuning iterations.
% \vspace{\baselineskip}

\section{Future Work}
% \subsection{Policy Expansion}
% Scaling this workflow to more policies (beyond Non-Family Safe) presents challenges.  Running expensive LLMs for every advertiser may become infeasible, necessitating higher-throughput, advertiser-level models trained on LLM outputs. 

% \subsection{Accuracy Improvements}
The method of iteratively tuning the prompt described above showed great promise for improving model performance. We plan to formalize this process such that we can categorize errors caught during re-review of good advertisers and adjust the prompt accordingly. 

% \subsection{Other Use Cases} 
In addition to providing the policy classification, the LLM is also prompted to give a summary of the advertiser and their products and services. We plan to experiment with providing such information to reviewers as \textit{hints} and look at the impact on reviewer performance (e.g. review speed, reliability, consistency, etc.).

\bibliographystyle{unsrt}
\bibliographystyle{ACM-Reference-Format}
\bibliography{sample-base}

\end{document}